\documentclass{article}
\usepackage[preprint]{spconfa4}
\usepackage{amsmath,graphicx}
\usepackage[nolist]{acronym}
\usepackage{amssymb}
\newcommand\norm[1]{\lVert#1\rVert}
\DeclareMathOperator*{\argmax}{arg\,max}
\newcommand*\bigstrut{%
  \vrule height\baselineskip depth.3\baselineskip width 0pt\relax}
\usepackage{cite}  
\usepackage{caption}
\usepackage{subcaption}
\usepackage{wrapfig}
\usepackage{environ}         
\usepackage{etoolbox}  
\usepackage{hyperref}
\usepackage{graphicx}        

\newlength{\myl}
\let\origequation=\equation
\let\origendequation=\endequation

\RenewEnviron{equation}{
  \settowidth{\myl}{$\BODY$}                       
  \origequation
  \ifdimcomp{\the\linewidth}{>}{\the\myl}
  {\ensuremath{\BODY}}                             
  {\resizebox{\linewidth}{!}{\ensuremath{\BODY}}}  
  \origendequation
}
\usepackage [english]{babel}
\usepackage [autostyle, english = american]{csquotes}
\MakeOuterQuote{"}

\title{BINAURAL SPEECH ENHANCEMENT USING STOI-OPTIMAL MASKS}
\name{Vikas Tokala, Mike Brookes, Patrick A. Naylor \thanks{This work was supported by funding from the European Union’s Horizon 2020 research and innovation programme under the Marie Skłodowska-Curie grant agreement No 956369 and the UK Engineering and Physical Sciences Research Council [grant number EP/S035842/1]}}
\address{Department of Electrical and Electronic Engineering, Imperial College London \\
\{v.tokala, mike.brookes, p.naylor\}@imperial.ac.uk} 
\begin{acronym}
\acro{SNR}{Signal-to-Noise Ratio}
\acro{SOBM}{STOI-optimal Binary Mask}
\acro{STOI}{Short-Time Objective Intelligibility}
\acro{WSTOI}{Weighted STOI}
\acro{HSWOBM}{High-resolution Stochastic WSTOI-optimal Binary Mask}
\acro{STFT}{Short Time Fourier Transform}
\acro{ILD}{Interaural Level Differences}
\acro{IPD}{Interaural Phase Differences}
\acro{ITD}{Interaural Time Differences}
\acro{LSA}{Log Spectral Amplitude}
\acro{OM-LSA}{Optimally-modified Log Spectral Amplitude}
\acro{DNN}{Deep Neural Network}
\acro{ERB}{Equivalent Rectangular Bandwidth}
\acro{DFT}{Discrete Fourier Transform}
\acro{VSSNR}{Voiced-Speech-plus-Noise to Noise Ratio}
\acro{SPP}{Speech Presence Probability}
\acro{HRIRs}{Head Related Impulse Response}
\acro{MBSTOI}{Modified Binaural STOI}
\acro{ReLU}{Rectified Linear Unit}
\acro{SNR}{Signal-to-Noise Ratio}
\acro{ISTFT}{Inverse STFT}
\acro{TF}{Time-Frequency}
\end{acronym}
\begin{document}

\maketitle
\begin{abstract}
 STOI-optimal masking has been previously proposed and developed for single-channel speech enhancement. In this paper, we consider the extension to the task of binaural speech enhancement in which the spatial information is known to be important to speech understanding and therefore should be preserved by the enhancement processing. Masks are estimated for each of the binaural channels individually and a `better-ear listening' mask is computed by choosing the maximum of the two masks. The estimated mask is used to supply probability information about the speech presence in each time-frequency bin to an \ac{OM-LSA} enhancer. We show that using the proposed method for binaural signals with a directional noise not only improves the SNR of the noisy signal but also preserves the binaural cues and intelligibility.
\end{abstract}
\vspace{-0.5cm}
\section{Introduction}
\label{sec:intro}
 For binaural signals, along with enhancement of \ac{SNR} and intelligibility, binaural cues must also be preserved. Studies have shown that preserving the binaural cues is helpful for sound localization and speech intelligibility in noisy environments due to \textit{binaural unmasking} \cite{Beutelmann2006,Lavandier2010a}. \ac{ILD} and \ac{ITD} are particularly helpful in localizing sound, dereverberation, improving intelligibility and boosting the perceived loudness \cite{blauert1997,hawley2004}. Binaural speech enhancement methods using beamformers \cite{Lotter2006,Hadad2015a}, multichannel Wiener filters \cite{Hadad2015,Doclo2006} and mask informed enhancement methods \cite{Moore2018b,Green2022} were previously proposed. 

Mask-based speech enhancement has been well studied for monoaural signals and has been shown to improve the \ac{SNR} and intelligibility \cite{Gonzalez2014a, Li2008}. The \ac{SOBM}, designed to optimize \ac{STOI}, was introduced in \cite{Lightburn2015} for monoaural signals. In this paper we propose a method for extending the monoaural mask-assisted enhancement to binaural speech by using \ac{HSWOBM} version of the \ac{SOBM}, which optimizes \ac{WSTOI}\cite{Lightburn2016} as our training target to estimate a continuous valued mask. Directly applying the \ac{TF} mask as a gain in the \ac{STFT} domain produces significant artefacts and so in \cite{Lightburn2017}, the \ac{OM-LSA} is proposed as an alternative method of applying the TF mask which improves the perceptual quality of the enhanced speech. 

The paper is organized with Sec.~\ref{sec:method} introducing the \ac{STOI} and \ac{WSTOI} metrics, \ac{STOI}-optimal mask based speech enhancement and the use of \ac{OM-LSA} and \ac{SPP} for mask application. Section~\ref{sec:simu} outlines the structure of the experiments followed by results and discussion in Sec.~\ref{sec:Results} and Sec.~\ref{sec:conclusions} draws the conclusions.

\section{Mask-based Speech Enhancement}
\label{sec:method}
\vspace{-.1cm}
\subsection{Overview of STOI and WSTOI}\label{sec:stoi}
\ac{STOI} is an intrusive intelligibility metric based on the correlation between the spectral envelopes of clean and degraded versions of the speech \cite{Taal2011}. To compute \ac{STOI}, these signals are converted into the \ac{STFT} domain using a 50\%-overlapping Hanning window of length $25.6$~ms and this results in the clean and degraded \ac{STFT} signals, $X(k,m)$ and $Y(k,m)$, with $(k,m)$ the frequency and time frame indices. $X(k,m)$ and $Y(k,m)$ are combined into $J=15$ third-octave bands by calculating the amplitudes of the TF cells denoted by $X_j(m)$ and $\tilde{Y}_j(m)$ where $j=1,..,J$ and $\widetilde{[\cdot]}$ indicates amplitude clipping to limit the impact of frames having low speech energy. The modulation vector is defined as $\mathbf{x}_{j,m}=[X_j(m-M+~1),X_j(m-M+2),..,X_j(m)]^T$, where $M=30$. The modulation vectors for clean and degraded signals, denoted by $\mathbf{x}_{j,m}$ and $\mathbf{y}_{j,m}$, are computed by performing a correlation between clean and degraded speech vectors of duration $(25.6 \times 30)/2=384$ ms. The clipped TF cell amplitudes, denoted by $\widetilde{[\cdot]}$, are determined as
\vspace{-0.2cm}
\begin{equation}\label{clipmod}
    \widetilde{Y}_j(m) = \mathrm{min}  \left( Y_j(m), \lambda \frac{\norm{\mathbf{y}_{j,m}}}{\norm{\mathbf{x}_{j,m}}} X_j(m) \right)
\end{equation}
\noindent where $\lambda = 6.623$ and $\norm{\cdot}$ denotes the Euclidean norm. The corresponding clipped modulation vector is $\mathbf{\tilde{y}}_{j,m}$.  The \ac{STOI} contribution of each TF cell $(j,m)$ is then given by
\begin{equation}
  d(\mathbf{x}_{j,m},\mathbf{\tilde{y}}_{j,m}) \triangleq \frac{(\mathbf{x}_{j,m} - \bar{x}_{j,m})^T \mathbf{\tilde{y}}_{j,m}}{\norm{\mathbf{x}_{j,m}- \bar{x}_{j,m}} \norm{\mathbf{\tilde{y}}_{j,m}- \bar{\tilde{y}}_{j,m}} },
\end{equation}

\noindent where $\bar{x}_{j,m}$ and $\bar{\tilde{y}}_{j,m}$ denote the mean of the vectors $\mathbf{x}_{j,m}$ and $\mathbf{\tilde{y}}_{j,m}$, respectively. The overall \ac{STOI} metric is computed by averaging the contributions of the TF cells over all bands, $j$, and frames, $m$.  In \cite{Lightburn2016}, the authors propose a modified version of \ac{STOI}, where each TF cell contribution for the final \ac{STOI} score is weighted by  estimated intelligibility content of the TF cell. Both metrics compare a clean speech signal with a degraded speech signal. Also, for computing the \ac{WSTOI}, the correlation comparison is computed on individual \ac{STFT} frequency bins rather than third-octave bands. The weight $I_{j,m}$ is calculated from the mutual information estimated using linear prediction models and is optimized using a language model \cite{kneser1995}. The \ac{WSTOI} score is given by
\begin{equation}\label{WSTOI}
 \textrm{WSTOI} = \frac{1}{\sum_{j,m} I_{j,m}} \sum_{j,m} I_{j,m}d(\mathbf{x}_{j,m}, \mathbf{\tilde{y}}_{j,m}).
\end{equation}

\subsection{Proposed Binaural Mask-based Enhancement}
 \begin{figure}
    \centering
    \includegraphics[scale=0.4]{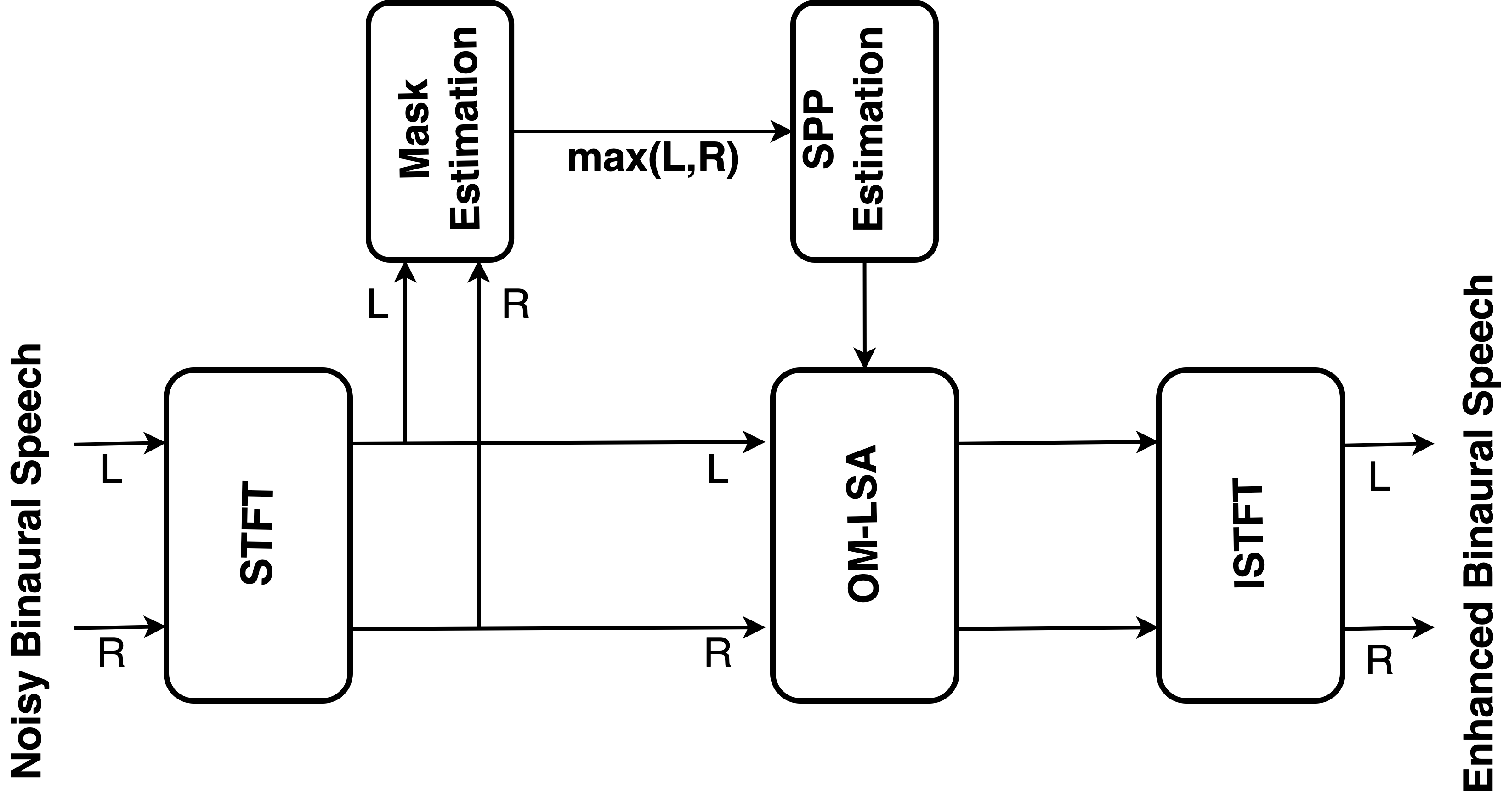}
     \vspace{-0.5cm}
    \caption{Block diagram of the proposed method.}
    \label{fig:BDBSWOBM}
    \vspace{-0.5cm}
\end{figure}
A block diagram of the proposed method is shown in Fig~\ref{fig:BDBSWOBM}. The L and R labels indicate the left and right channels of a binaural signal. The left and right channels are processed in parallel for all stages of the algorithm. The noisy binaural signal is transformed into the \ac{STFT} domain for TF-based processing. From the \ac{STFT} domain signals, 3 feature subsets as defined in Sec.~\ref{feat} are extracted once per \ac{STFT} frame and the same features are used for the training and prediction stages of the algorithm. \ac{STOI} optimal masks correlate to intelligible speech in the signal and \ac{SPP} is computed from the estimated continuous-valued mask and then supplied to the \ac{OM-LSA}. The output signal of the \ac{OM-LSA} is transformed back into to the time-domain by applying an \ac{ISTFT}.
\vspace{-0.3cm}
\subsubsection{Mask Estimation Features}\label{feat}
The feature extraction and mask estimation follows the method from \cite{Lightburn2020}. The first step in feature extraction for mask estimation is to normalize the noisy speech active level to 0 dB using the ITU P.56 objective speech active level \cite{ITU_T_P56_TRa} to make the algorithm level independent. Each of the three feature subsets has $\Psi$ features, giving $3\Psi$ features in total.

The first feature subset is constructed from the TF gains estimated by the Log-MMSE speech enhancement algorithm \cite{Ephraim1985,Brookes1997}, that minimizes the mean-squared error in the log-spectral amplitudes. The TF gains are expected to be high when speech is present and low otherwise. As this algorithm uses a noise estimator \cite{Gerkmann2012}, these features are expected to help generalize the mask estimator to unseen noise types. The gain function $G(k,m)$ from \cite{Ephraim1985} satisfies $G(k,m) |Y(k,m)| = \mathrm{exp}\{\mathbb{E}[\textrm{log}|X(k,m)| \; |Y(k,m)|]\}$ where, $\mathbb{E[.]}$ is the expectation operator. The first subset of features, $\boldsymbol{\mu}_{m}^{(1)} = \left[\mu_{1,m}^{(1)},...,\mu_{\Psi,m}^{(1)}\right]^T$, is a $\Psi \times 1$ vector found by averaging $G(k,m)$ in $\Psi=30$ triangular windows, $w_i(k)$ for $i=1,..,\Psi$ equally spaced on \ac{ERB} scale centre frequencies and then computing the natural logarithm. For frame $m$, the feature subset is given by
\vspace{-0.2cm}
\begin{equation}
  \mu_{i,m}^{(1)} = \textrm{ln}\left \{  \frac{\sum_{k=0}^{K/2} w_i(k) G(k,m)}{\sum_{k=0}^{K/2} w_i(k)} \right\} \textrm{for} \; i=1,...,\Psi
\end{equation}
\noindent where $K$ is the \ac{DFT} length. The second feature subset, denoted by $\boldsymbol{\mu}_{m}^{(2)} = \left[\mu_{1,m}^{(2)},...,\mu_{\Psi,m}^{(2)}\right]^T$, is the estimate of the level-normalized enhanced speech amplitude in each frequency band and is obtained by multiplying the gains from the first feature subset with the noisy speech. The third feature subset $\boldsymbol{\mu}_{m}^{(3)} = \left[\mu_{1,m}^{(3)},...,\mu_{\Psi,m}^{(3)}\right]^T$ is an estimate of the local \ac{VSSNR} in the TF regions and is obtained using the PEFAC algorithm \cite{Gonzalez2014}. The fundamental frequency of the speech in each frame is computed and then \ac{VSSNR} is estimated within the $\Psi$ frequency bands by comparing the energy at harmonics of the fundamental frequency with the energy halfway between consecutive harmonics. The complete feature set for frame $m$ is obtained by concatenating the three subsets to obtain the feature set $ \boldsymbol{\mu}_m = \left[\boldsymbol{\mu}_{m}^{(1)T}, \boldsymbol{\mu}_{m}^{(2)T}, \boldsymbol{\mu}_{m}^{(3)T} \right]^T$. 
\vspace{-0.3cm}
\subsubsection{Target Mask and \ac{DNN}} \label{mask}
The difference between \ac{SOBM} and \ac{HSWOBM} is that the latter is designed to optimize \ac{WSTOI} rather than \ac{STOI} and is computed over a higher number of frequency bands (129 instead of 15) to increase the resolution of the mask. In the proposed method, \ac{HSWOBM}s are used as target masks and are computed using the three-stage dynamic programming technique described in \cite{Lightburn2015} which consists of carrying out three passes of estimation with a pruning strategy to compute the target masks to optimise the \ac{WSTOI} from \eqref{WSTOI}. Directional white Gaussian noise at 0 dB SNR is added to binaural speech signals to generate the training data and computation of target masks. The target masks are computed for the left and right channels individually based on the speech content in each channel.  The \ac{HSWOBM} is computed by forming a masked signal $Z_j(m) = B_j(m)Y_j(m)$ where $B_j(m) \in \{0,1\}$. The clipped modulation masked vector $\mathbf{\tilde{z}}_{j,m}$ is then computed analogous to \eqref{clipmod} to limit the impact of frames having low speech energy. $B_j$ is numerically optimized using least squares method separately in each band, $j$, by computing
\vspace{-0.25cm}
\begin{equation*}
B_j(m) = \argmax_{\{B_j(m):m=1,...,N\}}\left(\sum_{m=1}^N I_{j,m} \mathbb{E}[  d(\mathbf{x}_{j,m},\mathbf{\tilde{z}}_{j,m}) ] \right)
\end{equation*}

\noindent where $N$ is the number of frames and
\vspace{-0.2cm}
\begin{equation}
   \mathbb{E}[d(\mathbf{x}_{j,m},\mathbf{\tilde{z}}_{j,m}) ] \approx \frac{(\mathbf{x}_{j,m} - \bar{x}_{j,m})^T \;\mathbb{E}[ \mathbf{\tilde{z}}_{j,m}]}{\norm{\mathbf{x}_{j,m}- \bar{x}_{j,m}} \; \sqrt{\bigstrut \mathbb{E}[ \norm{\mathbf{\tilde{z}}_{j,m}- \bar{\tilde{z}}_{j,m}}^2}] }.
\end{equation}
A feed-forward \ac{DNN} is trained to estimate a continuous mask from the features extracted from the noisy binaural signals by using the \ac{HSWOBM} target masks estimated by dynamic programming \cite{Lightburn2015,Lightburn2020}. The \ac{DNN} consists of 4 layers of `dense' or `fully-connected' layers each with 500 hidden neurons. \ac{ReLU} is used as the activation function for all the layers except the output layer where the Sigmoid activation function is used. Dropout layers are placed between the layers with a dropout factor of 20\%. A weighted mean square error is used as the cost function for the \ac{DNN}. The cost function, $\mathcal{J}$, is given by
\vspace{-0.2cm}
\begin{equation}\label{costfunc}
  \mathcal{J} = \frac{1}{\Delta}\frac{\sum_{p=1}^\Delta \phi_p (\mathrm{c}_p - \zeta_p)^2}{ \sum_{p=1}^\Delta \phi_p}
\end{equation}
where $c_p$ is the output of the \ac{DNN}, $\Delta$ is the number of pairs of feature vectors and corresponding mask value pairs available for training, $\zeta_p$ is the training target and $\phi_p$ is the corresponding value of the weight. The weights are computed from the \ac{WSTOI} sensitivity and band importance weighting obtained from the \ac{WSTOI} metric \cite{Lightburn2016}.

\vspace{-0.25cm}
\subsubsection{Better-ear Processing} 
 Studies \cite{blauert1997, Green2022} have shown that human listeners are capable of better ear listening, which is focusing on the speech signal from the ear which has higher SNR and intelligibility. We adopt a similar approach in our method for the selection of the mask values or the gains in each TF bin. Choosing the maximum gain value from the two estimated masks for the left and right channels is equivalent to selecting the gain associated with lower noise levels and higher speech presence from the \ac{SPP} estimation. In order to preserve the spatial cues of the binaural signal, the same gain needs to be applied to both the channels. Hence, in our method, we use the gains selected from the above method to compute a single \ac{SPP} to input to the \ac{OM-LSA} enhancer for both the channels. 
\vspace{-0.25cm}
\subsubsection{Optimally Modified-LSA}
Mask-application follows the method in \cite{Lightburn2017}, that is shown to improve the perceptual quality of masked speech. From the computed \ac{HSWOBM}, an intermediate  continuous-valued mask is estimated and, instead of applying this mask as a TF domain gain, it is used to supply the probability of speech presence to a speech enhancer \cite{Cohen2002a} that minimizes the expected error in the \ac{LSA}. Let us consider the hypothesis $H_1(k,m)$ in the $k^{th}$ bin where speech is present and $H_0(k,m)$ where speech is absent. Considering the gain $G(k,m)$ to be larger than a threshold $G_{min}$ when the speech is absent, it is shown in \cite{Cohen2002a} that

\begin{equation}
    G(k,m) = \{ G_{H_1}(k,m)\}^{p(k,m)} \; G_{min}^{1-p(k,m)}
\end{equation}

\noindent where $G_{H_1}(k,m)$ is the gain under the hypothesis $H_1(k,m)$ and $p(k,m) \triangleq P(H_1(k,m) | Y(k,m))$ is the conditional \ac{SPP} and is given by
\vspace{-0.2cm}
\begin{equation*}
    p(k,m) = \left \{ 1 + \frac{q(k,m)}{1-q(k,m)} (1+ \xi (k,m)) \mathrm{exp}(-v(k,m)) \right\}^{-1}
\end{equation*}
\noindent where $q(k,m)~\triangleq~P(H_0(k,m))$ is the \textit{a priori} probability of speech absence in the hypothesis $H_0(k,m)$, $v(k,m)$ is the \textit{a posteriori} \ac{SNR} and $\xi (k,m)$ is the \textit{a priori} \ac{SNR} . In \cite{Lightburn2017} it is shown that, by using the gain value of the mask to control the speech probability and using this to enhance the speech using \ac{OM-LSA}, there is a soft imposition of the spectro-temporal modulations required to preserve the intelligibility of the enhanced speech. From the estimated better-ear continuous mask, the \ac{SPP} which will be used to enhance both the channels is computed and supplied to the \ac{OM-LSA}. The enhanced signals are then transformed into the time domain from the TF domain by performing the inverse \ac{STFT}.

\begin{figure*}[!h]
     \centering
     \begin{subfigure}[b]{0.33\textwidth}
         \centering
         \includegraphics[width=0.95\textwidth]{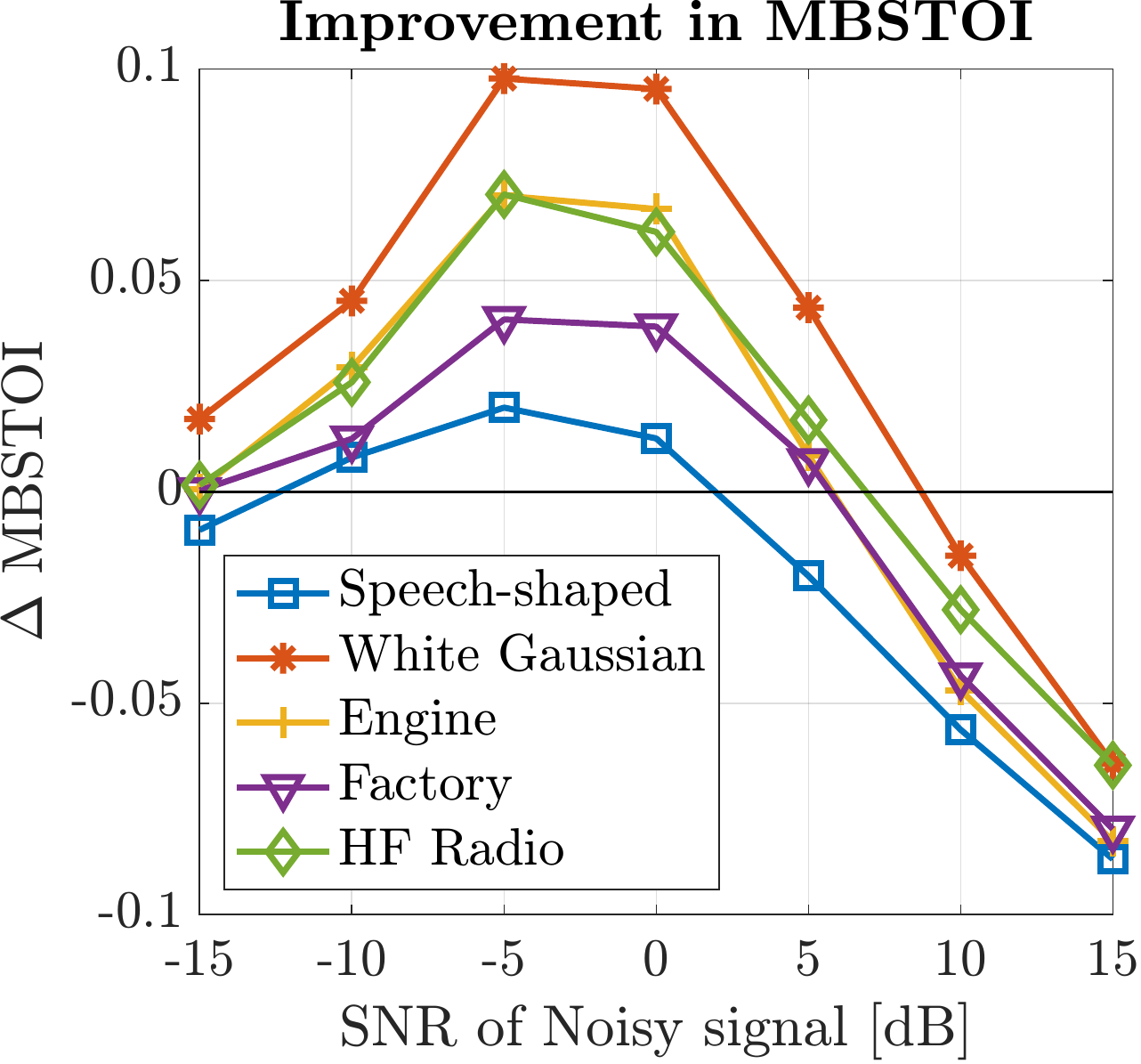} 
         \caption{}
         \label{fig:mbstoi}
     \end{subfigure}
     \hfill
     \begin{subfigure}[b]{0.33\textwidth}
         \centering
         \includegraphics[width=0.95\textwidth]{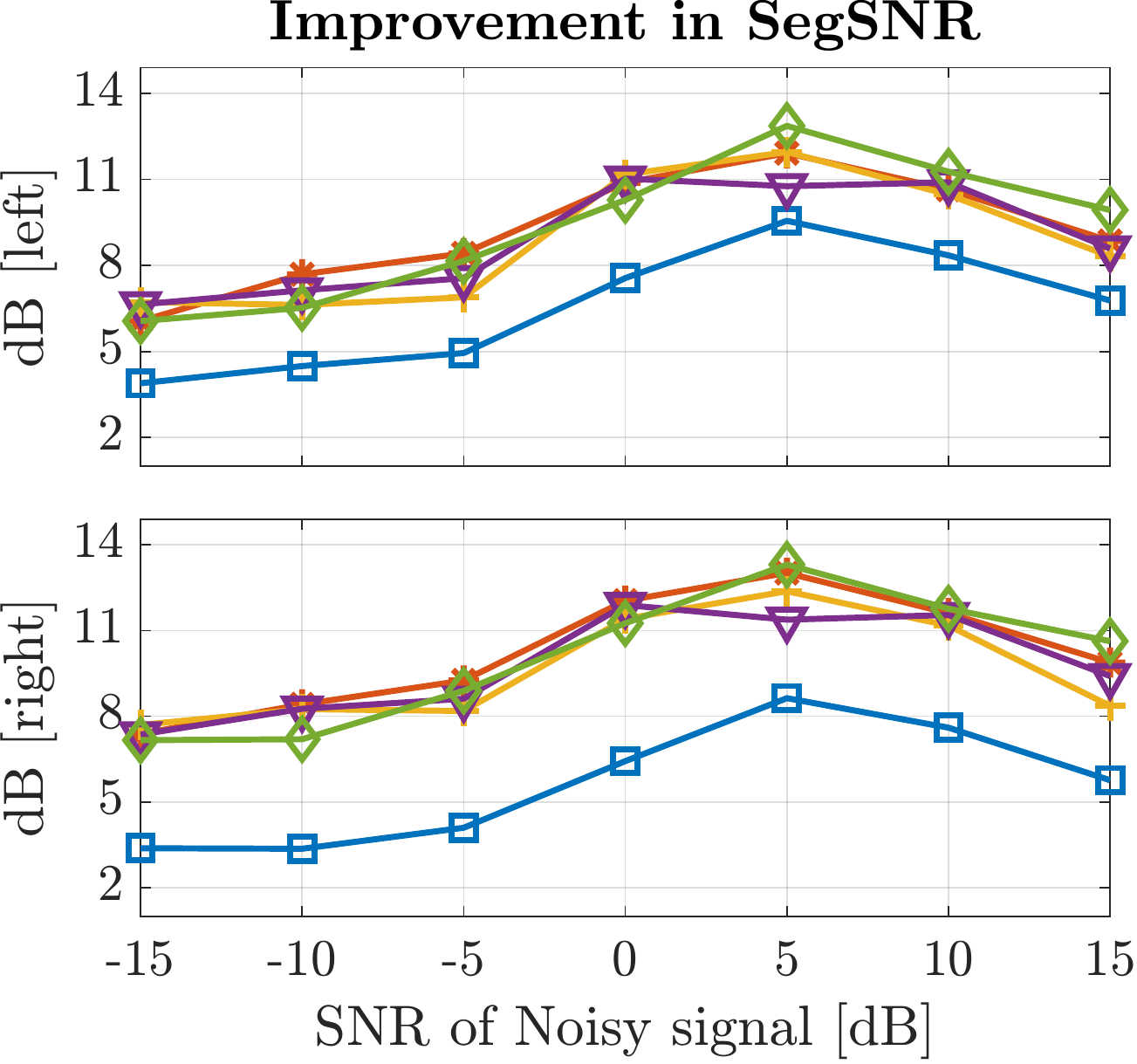}
         \caption{}
         \label{fig:fssnr}
     \end{subfigure}
     \hfill
     \begin{subfigure}[b]{0.33\textwidth}
         \centering
         \includegraphics[width=0.98\textwidth]{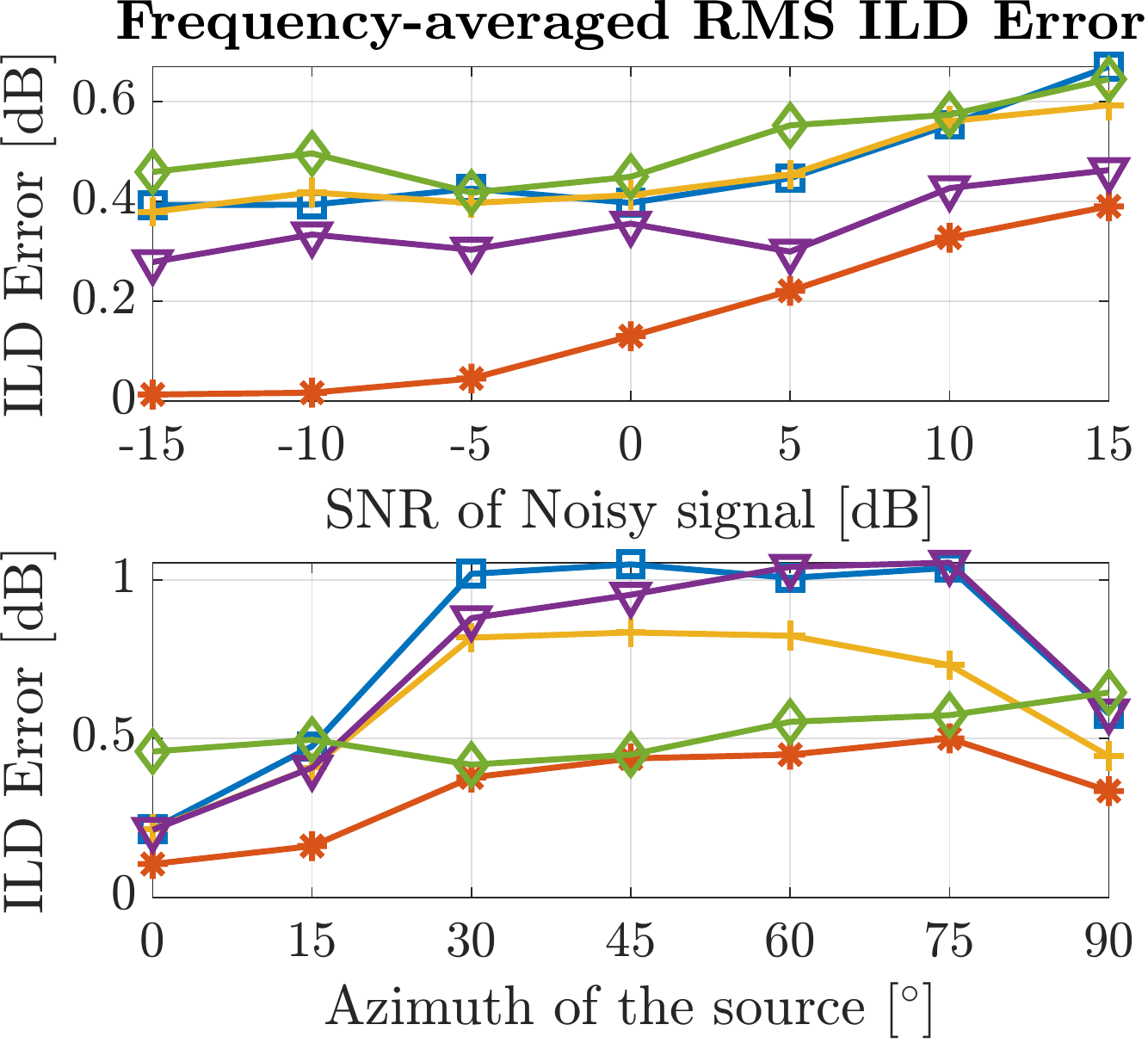}
         \caption{}
         \label{fig:ild}
     \end{subfigure}
      \vspace{-0.7cm}
        \caption{(a) Improvement in \ac{MBSTOI} score, (b) improvement in frequency weighted segmental \ac{SNR} for both channels and (c)(upper) frequency averaged RMS ILD error with different \ac{SNR}s, for a source placed 80 cm away at $30^\circ$ azimuth and noise at $0^\circ$ azimuth (c)(lower) frequency averaged RMS ILD error for source at 80 cm with 0 dB SNR for different azimuths of the source and the noise at $0^\circ$ azimuth.  }
        \vspace{-0.5cm}
\end{figure*}
\vspace{-0.2cm}
\section{Simulation Experiments}
\vspace{-0.1cm}
\label{sec:simu}
 Simulations were performed using speech utterances from TIMIT \cite{Garofolo1993} and noise signals from NOISEX-92~\cite{Varga1993} and all signals were resampled to 10 kHz. To simulate binaural signals and directional noise, the \ac{HRIRs} from \cite{Kayser2009} were used. To generate the training target \ac{HSWOBM} masks, 400 speech signals from TIMIT and the anechoic in-ear \ac{HRIRs} were used. The source  azimuth was randomly selected between $-90^{\circ}$ and $+90^{\circ}$ with a resolution of $5^{\circ}$ and with elevation and distance fixed at 80 cm and $0^\circ$. The target masks for the left and right channels were then computed as described in Sec.~\ref{mask}. A feed-forward \ac{DNN} was designed in Python using Tensorflow. Around 2 million trainable parameters were generated from the three extracted feature sets. The \ac{DNN} was trained to estimate a continuous mask to optimize the \ac{WSTOI} metric in each channel by minimizing the cost function in \eqref{costfunc}. The generated trainable data was split into 70\% for training and 30\% for validation while training the model. For evaluation, 30 unseen male and female speech utterances randomly selected from TIMIT test dataset,  and 5 types of noise signals from NOISEX-92, and spatialized using \ac{HRIRs} \cite{Kayser2009} were used to generate the test data. 

Experiments with different SNRs were performed by adding a normalized directional noise at $0^\circ$ azimuth to a normalized binaural speech signal at SNRs in the range of -15~dB to +15~dB in steps of 5~dB. The A-weighted frequency segmental SNR \cite{Hu2006a, Brookes1997} is measured to show the noise reduction, \ac{MBSTOI} \cite{Andersen2018} to show the binaural intelligibility improvement and the \ac{ILD} error to quantify the preservation of binaural cues. The ILD for binaural signals is computed in each TF bin as $\textrm{ILD}_{(k,m)} = 20 \textrm{log}_{10}(|Y_L(k,m)|/|Y_R(k,m)|) $, where $Y_L$ and $Y_R$ are the left and right channel \ac{STFT} domain signals. The RMS \ac{ILD} error is computed by averaging the error over all the frequencies and for \ac{SNR}s in the range of -15~dB to +15~dB in steps of 5~dB of received binaural speech and at azimuths 0 to $90^\circ$ in steps of $15^\circ$. The RMS ILD error was averaged over frequency because in our experiments no significant changes in the error were observed over frequencies. In both the experiments to compute the RMS ILD error, the directional noise was placed at $0^\circ$ azimuth. The proposed method does not alter the phase of the signal while processing and therefore \ac{ITD}s of the enhanced signal will be the same as the signal before processing. 

\vspace{-0.3cm}
\section{Results}
\vspace{-0.1cm}
\label{sec:Results}
  Figure~\ref{fig:mbstoi} shows the improvement in \ac{MBSTOI} and maximum improvement was observed between -5 to 0
  ~dB SNR for all noise cases. At very noisy \ac{SNR}s of under -10~dB, extraction of speech features from the signal to estimate an accurate mask becomes difficult, which explains lower improvements. However, in Fig.~\ref{fig:fssnr}, which shows the improvement in segmental \ac{SNR}, even under very noisy cases of -15~dB SNR where the speech information in the extracted features is low, the method was able to provide at least 3~dB SNR improvement for all the noise cases and as the signal SNR improved, SNR improvements greater than 10 dB were also observed. For \ac{SNR}s higher than 5~dB, there is negative improvement in \ac{MBSTOI}, as the input binaural speech signal has a higher \ac{MBSTOI} score before processing and applying mask based enhancement on these signals results in reducing the score. As all the signals above 5~dB SNR had a \ac{MBSTOI} score above 0.7 after processing, the overall intelligibility of speech signals did not deteriorate to unintelligible levels. From Fig.~\ref{fig:fssnr}, it can seen that a similar improvement in segmental SNR can be observed for all the noise cases with speech shaped noise being the exception and also from Fig.~\ref{fig:mbstoi}, lowest improvement for \ac{MBSTOI} can be seen for speech shaped noise. Although speech shaped noise is a stationary noise, as it has most of its energy located in the TF bins shared with speech, thus impacting the estimation of \ac{SPP} and hence the performance. For the experiments with multiple azimuth angles, the source was simulated at azimuths from $0^\circ$ to $90^\circ$ since the other half of the frontal plane in the \ac{HRIRs} database \cite{Kayser2009}  is a reflection of the former. Figure~\ref{fig:ild} shows the frequency averaged RMS \ac{ILD} error observed with the method for input SNRs of the noisy signal and for different azimuth angles of the source. For all the noise cases, SNRs and azimuths the RMS \ac{ILD} error observed is under 1~dB in the frontal plane. For all azimuths and SNRs, the proposed method performed the best for white Gaussian noise having the least error among all the noise cases. This low RMS \ac{ILD} error after processing quantifies the binaural cue preservation of the method.

 \vspace{-0.5cm}
 \section{Conclusions}\label{sec:conclusions}
 \vspace{-0.2cm}
 We have presented a new approach using \ac{STOI}-optimal masking to enhance noisy binaural speech signals that not only improves the SNR but also preserves the interaural cues and intelligibility. Instead of applying the TF masks as a gain function, \ac{SPP} is calculated and used by an \ac{OM-LSA} speech enhancer. The proposed method is able to provide significant improvements in frequency weighted SNR, \ac{MBSTOI} score and minimal distortion of interaural cues.

\bibliographystyle{IEEEtran}
\bibliography{sapstrings,sapref}

\end{document}